\documentclass[
 aps,
 prd,
 reprint,
 amsmath,amssymb,
 aps,floatfix,
]{revtex4-2}

\usepackage{graphicx}
\usepackage{dcolumn}
\usepackage{bm}
\usepackage{mathtools}

\usepackage{braket}

\RequirePackage{color}
\definecolor{dkgreen}{rgb}{0.2,0.7,0.4}
\definecolor{dkblue}{rgb}{0.2,0.2,0.7}
\definecolor{dkred}{rgb}{0.8,0,0}
\definecolor{dkgreen}{rgb}{0.2,0.8,0.4}

\begin{document}



\title{Quantum Many-Body Scars for Arbitrary Integer Spin in 2+1D Abelian Gauge Theories}

\author{Thea Budde}
\author{Marina Krsti{c} Marinkovi{c}}
\author{Joao C. Pinto Barros}
\affiliation{Institut f\"{u}r Theoretische Physik, ETH Z\"{u}rich,
Wolfgang-Pauli-Str. 27, 8093 Z\"{u}rich, Switzerland}

\date{\today}

\begin{abstract}

The existence of Quantum Many-Body Scars, 
which prevents thermalization from certain initial states after a long time, has been established across different quantum many-body systems. These 
include gauge theories corresponding to spin-1/2 quantum link models. 
Establishing quantum scars in gauge theories with high spin is not accessible with existing numerical methods, which rely on exact diagonalization. 
We systematically identify scars for pure gauge theories with arbitrarily large integer spin $S$ in $2+1$D, where the electric field is restricted to $2S+1$ states per link. Through an explicit analytic construction, we show that the presence of scars is widespread in $2+1$D gauge theories for arbitrary integer spin. We confirm these findings numerically for small truncated spin and $S=1$ quantum link models. 
Our analytic construction establishes the presence of scars far beyond volumes and spins that can be probed with existing numerical methods and can guide quantum simulation experiments toward interesting non-equilibrium phenomena, inaccessible otherwise. 

\end{abstract}

\maketitle

\section{Introduction}
Evolving an isolated quantum system over long times might suggest that the unitary evolution of the system would prevent thermalization.
This puzzle is addressed by the Eigenstate Thermalization Hypothesis (ETH) for quantum many-body systems \cite{deutsch1991quantum,srednicki1994chaos,rigol2008thermalization, d2016quantum,deutsch2018eigenstate}. According to the ETH, for non-integrable systems, the high-energy states and the observables of interest will eventually converge to a description following equilibrium statistical mechanics, achieving thermal equilibrium. In contrast to the ETH, many-body localized systems defy this trend due to their emergent integrability~\cite{abanin2019colloquium}. Systems exhibiting Quantum Many-Body Scars (QMBS) offer a contrasting example~\cite{serbyn2021quantum,moudgalya2022quantum,chandran2023quantum}; they also evade the ETH, but only in an exponentially small fraction of states in the Hilbert space. 

The effect of QMBS was first observed in a Rydberg-atom quantum simulator, marked by persistent revivals for particular initial states, in contrast to the vast majority of other high-energy initial states~\cite{bernien2017probing,Bluvstein_2021}. In parallel with the exact construction of highly excited eigenstates in a non-integrable model \cite{moudgalya2018exact,moudgalya2018entanglement}, the experiment has sparked intense research in a variety of quantum many-body systems, where a \emph{weak} breaking of the ETH takes place \cite{Turner_2018,ho2019periodic,khemani2019signatures,choi2019emergent,schecter2019weak,mark2020eta,mark2020unified,su2023observation,halimeh2203robust,desaules2023weak,pakrouski2020many,pakrouski2021group,kolb2023stability}. In these systems, the presence of \emph{few} anomalous eigenstates can leave an imprint on thermalization. 
Among the special features of the anomalous states is that they are characterized by atypically low entanglement entropy, compared with other states arbitrarily close in the spectrum. In particular, QMBS may exhibit an area law of bipartite entanglement entropy. This means that the entanglement between a subsystem and its complement is proportional to the area of the boundary that divides them. This contradicts the ETH, which predicts that highly excited states have an entanglement that grows linearly with the volume of the subsystem.

The interplay between gauge symmetry and scarring phenomena is not fully understood. Gauge theories naturally lead to constraints, and specific constraints can give rise to QMBS, as observed in the PXP model~\cite{bernien2017probing,surace2020lattice, PhysRevLett.127.090602}. 
The constraints in gauge theories stem from local symmetries,
which segment the Hamiltonian into distinct sectors that do not mix under time evolution. 
Supplied by a condition on physical states, this leads to local constraints in the form of Gauss' law.
Beyond the PXP model, QMBS have also been observed in other gauge theories, such as the Abelian case in the presence of matter in $1+1$D \cite{desaules2023prominent}, pure gauge in $2+1$D \cite{banerjee2021quantum,biswas2022scars,sau2023sublattice} and non-Abelian in $1+1$D \cite{calajo2024quantum} and $2+1$D \cite{hayata2023string}. 
So far, the research on QMBS in Abelian pure gauge theories has exclusively focused on spin-$1/2$ Quantum Link Models (QLM) \cite{banerjee2021quantum,biswas2022scars,sau2023sublattice}, which also
exhibit other interesting phenomena such as crystalline confining phases or confining strings with
fractionalized electric flux stands \cite{banerjee2013crystalline,banerjee20132+,banerjee2018s,banerjee2022nematic}.
{This work demonstrates, for the first time, the existence of QMBS in link models with arbitrary integer spins.}

Addressing questions related to real-time dynamics 
falls outside the capabilities of conventional lattice gauge theory methods, as Monte Carlo simulations 
face inefficiencies due to severe sign problems (see e.g.~\cite{troyer2005computational,alexandru2022complex}).
Sign problems have been a driving force in the development of quantum simulations of gauge theories \cite{wiese2013ultracold,dalmonte2016lattice,Davoudi:2019bhy,banuls2020simulating,zohar2022quantum,halimeh2022stabilizing,di2023quantum,Bauer:2023qgm}. Formulations like QLMs are essential for preserving gauge symmetry while ensuring a finite Hilbert space for a finite lattice volume.
Recovering the original theory 
may take different routes~\cite{brower1999qcd,brower2004d,chandrasekharan1997quantum,Bhattacharya:2020gpm,zache2022toward,wiese2022quantum,halimeh2022achieving}.
Regardless of the chosen method, significant contributions to theories such as Quantum Chromodynamics (QCD) are still far off. Therefore, it is crucial to identify intriguing phenomena that can guide experimental efforts in this direction. The study of QMBS is a meritorious example, as they appear in simple gauge theories and probe fundamental aspects of quantum many-body theory.

Demonstrating the presence of QMBS for large Hilbert spaces per gauge link, beyond spin-$1/2$, serves several fundamental purposes: it directly addresses how widespread QMBS are across many-body systems; it reveals novel mechanisms for their formation, illuminating the role played by gauge symmetry; and it helps to guide quantum simulation experiments toward interesting questions before the complexity of QCD is achieved. 

In this work, we demonstrate the extensive presence of QMBS across $2+1$D $U(1)$ gauge theories without matter. Concretely, we explicitly construct mid-spectrum states that satisfy area law entanglement for arbitrary integer spin, specifically for a simply truncated Hilbert space per link. We further verify the existence of these states by numerically determining the system's eigenstates and calculating their entanglement and Shannon entropy for spin 1 and 2. 
Our findings unveil the presence of QMBS for single-leg ladders of the spin-1 QLM, which are more accessible to experiments than wider systems.
Furthermore, our analytical approach allows us to identify QMBS for arbitrary integer spin and system size, circumventing the limitations of existing numerical methods when applied to large dimensional Hilbert spaces. 

\section{Link models in $2+1$D}
We consider models on square $L_1\times L_2$ lattices with bosonic degrees of freedom living on the links
\begin{equation}
    \label{eq:H}
    H=\sum_{n}\left( U_{n1}^\dagger U_{n+\hat{1}2}^\dagger U_{n2} U_{n+\hat{2}1}+\mathrm{h.c.}\right)
    +V,
\end{equation}
where the indices $n\equiv\left(n_1,n_2\right)$ represent lattice sites, while the labels $i\in\left\{1,2\right\}$ represent the two directions. 
The first terms of the Hamiltonian are plaquette terms, constructed by acting on each of the four links of a plaquette with operators represented by $U$. The term $V$, which we will call generically \emph{potential}, will always be diagonal in the electric field $E_{ni}$. The commutation relations between these variables are $\left[E_{mi},U_{nj}\right]=U_{mi}\delta_{mn}\delta_{ij}$. Under these general conditions, the Hamiltonian has a set of local symmetries. Concretely, there is one generator of local gauge transformations $G_n$ per lattice site, which commutes with the Hamiltonian 
\begin{equation}
    \label{eq:Gn}
    G_n=E_{n1}+E_{n2}-E_{n-\hat{1}1}-E_{n-\hat{2}2},\quad \left[H,G_n\right]=0.
\end{equation}
We will use the electric field basis, denoted by $\left|\varepsilon\right>$ for a single link. The generators $G_n$ are diagonal in this basis. The Hilbert space breaks into many different sectors. We will focus on the physical sector characterized by the Gauss' law $G_n\left|\psi\right>=0$, for all sites, which acts as a constraint of the system.

The model also has two winding symmetries. Concretely, the observables 
$W_1=\sum_{n_1=0}^{L_1-1}E_{\left(n_1,m\right)2}$ and $W_2=\sum_{n_2=0}^{L_2-1}E_{\left(m,n_2\right)1}$ 
commute with the Hamiltonian. In the physical sector, their eigenvalues are independent of $m$. We focus on the zero-winding sector   $W_i\left|\psi\right>=0$ for both $i=1$ and $i=2$. This is the largest sector of the theory. There are further global symmetries that depend on the choice of the potential and are used to reduce the size of the Hilbert space in exact diagonalization calculations. They are described in the Supplementary Material \cite{supplementary}. 
We will be interested in two specific versions of this model that ensure that we will have a finite-dimensional Hilbert space per link.

In Quantum Link Models (QLM) \cite{horn1981finite,orland1990lattice,chandrasekharan1997quantum} $U_{ni}/U_{ni}^\dagger$ are spin raising/lowering operators. $U_{ni}$ and $U^\dagger_{ni}$ do not commute and act according to
$U\left|\varepsilon\right>\propto\sqrt{S\left(S+1\right)-\varepsilon\left(\varepsilon+1\right)}\ket{\varepsilon + 1}$. The electric field operators $E$ correspond to the $z$ component spin operators.

In Truncated Link Models (TLM) \cite{zohar2012simulating} (see also \cite{desaules2023prominent,popov2024non})
the electric field is simply truncated, meaning $U\left|S\right>=0$, $U^\dagger\left|-S\right>=0$ and $U\left|n\right>= \ket{n+1}$ otherwise. $U_{ni}$ and $U^\dagger_{ni}$ still do not commute, though the non-zero commutation relation is removed to the edge of the local Hilbert space (to the states $\ket{-S}$ and $\ket{S}$). With this construction, there will be $2S+1$ states per link. We generically refer to $S$ as the total spin, irrespective of the formulation we are using. The $S=1$ TLM is equivalent to the $S=1$ QLM, up to normalization of Hamiltonian parameters.

We will also write the Hamiltonian \eqref{eq:H} as
\begin{equation}
    \label{eq:Hdecomposition}
    H=K+V,
    \quad K=H^++H^-,\quad
    H^\pm=\sum_{n}H^\pm_n,
\end{equation}
where $H^-_n=U_{n1}^\dagger U_{n+\hat{1}2}^\dagger U_{n2} U_{n+\hat{2}1}$ and $H^+=\left(H^-\right)^\dagger$ are \emph{kinetic operators}. This notation follows the convention that we will adopt throughout the paper. ``Plaquette at site $n$'' refers to the plaquette for which $n$ is its lower left vertex.
In order to simplify the formulas we will also adopt the convention $\left(H^+\right)^{-m}\equiv \left(H^-\right)^{m}$. This should not be confused with the inverse of $\left(H^+\right)^m$, since these operators are not invertible.

To demonstrate the existence of scars in these models, one should show that they are non-integrable. It is expected that this is the case for pure gauge theories in $2+1$D, and we will further demonstrate it here for a single-leg ladder with $S=1$.

\subsection{Non-Integrability of the Spin-1 Ladder}

We consider a single row of plaquettes with spin $S=1$ links, open boundary conditions in both directions and the potential $V=\lambda\sum_{n_1=0}E_{\left(n_1,1\right)1}$. 
We show that his model is non-integrable by studying the statistics of the spectrum between adjacent energy levels. We follow the method introduced in \cite{oganesyan2007localization}, which was also applied to QLMs in \cite{banerjee2021quantum,biswas2022scars}.
This will be one of the many cases in which we demonstrate the existence of mid-spectrum low entropy states, establishing the presence of quantum many-body scars. While we do not demonstrate the non-integrability of all models considered, the result is expected to extend beyond this case.

\begin{figure}
    \centering
    \includegraphics[width=0.49\linewidth]{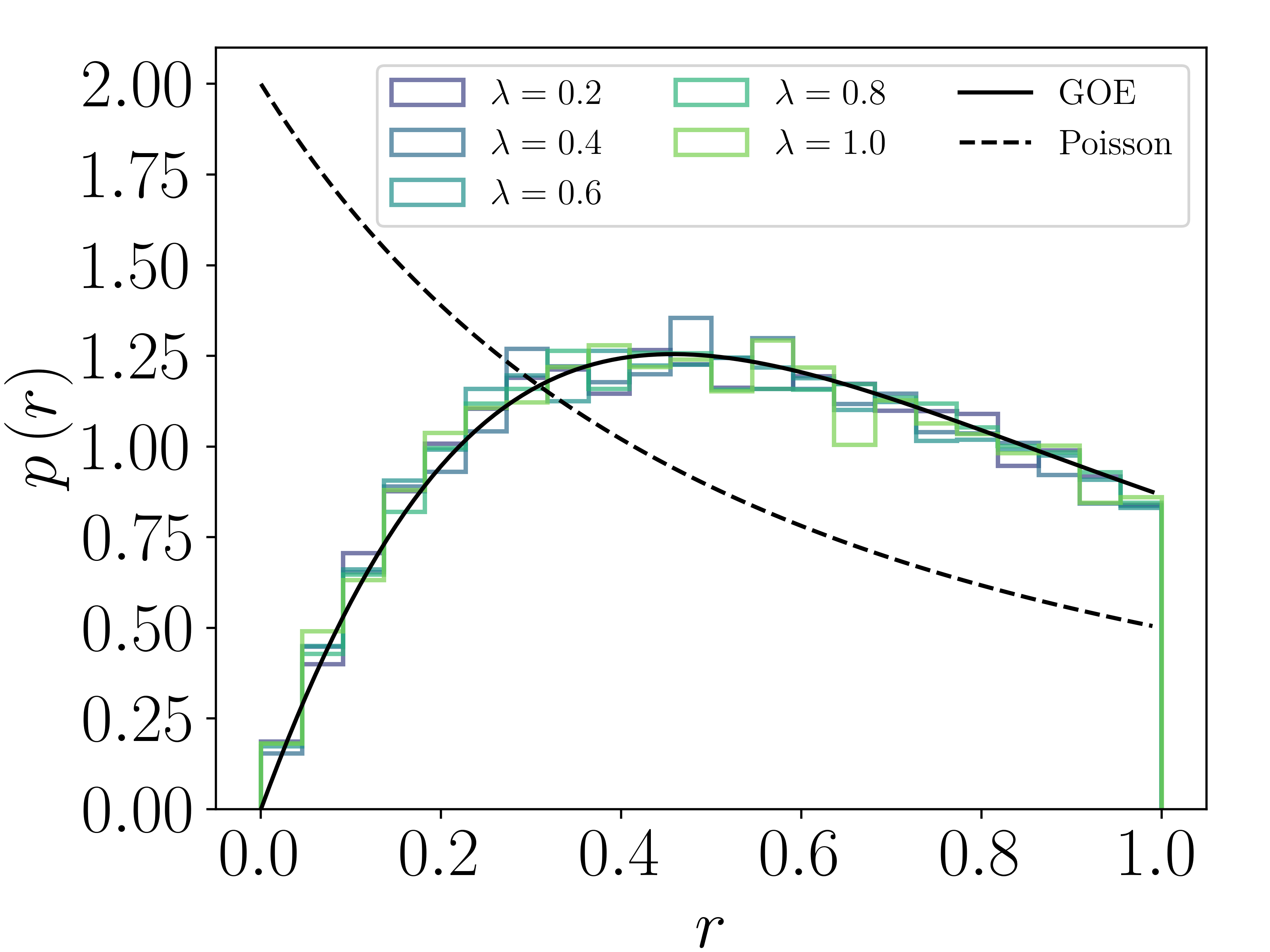}
    \includegraphics[width=0.49\linewidth]{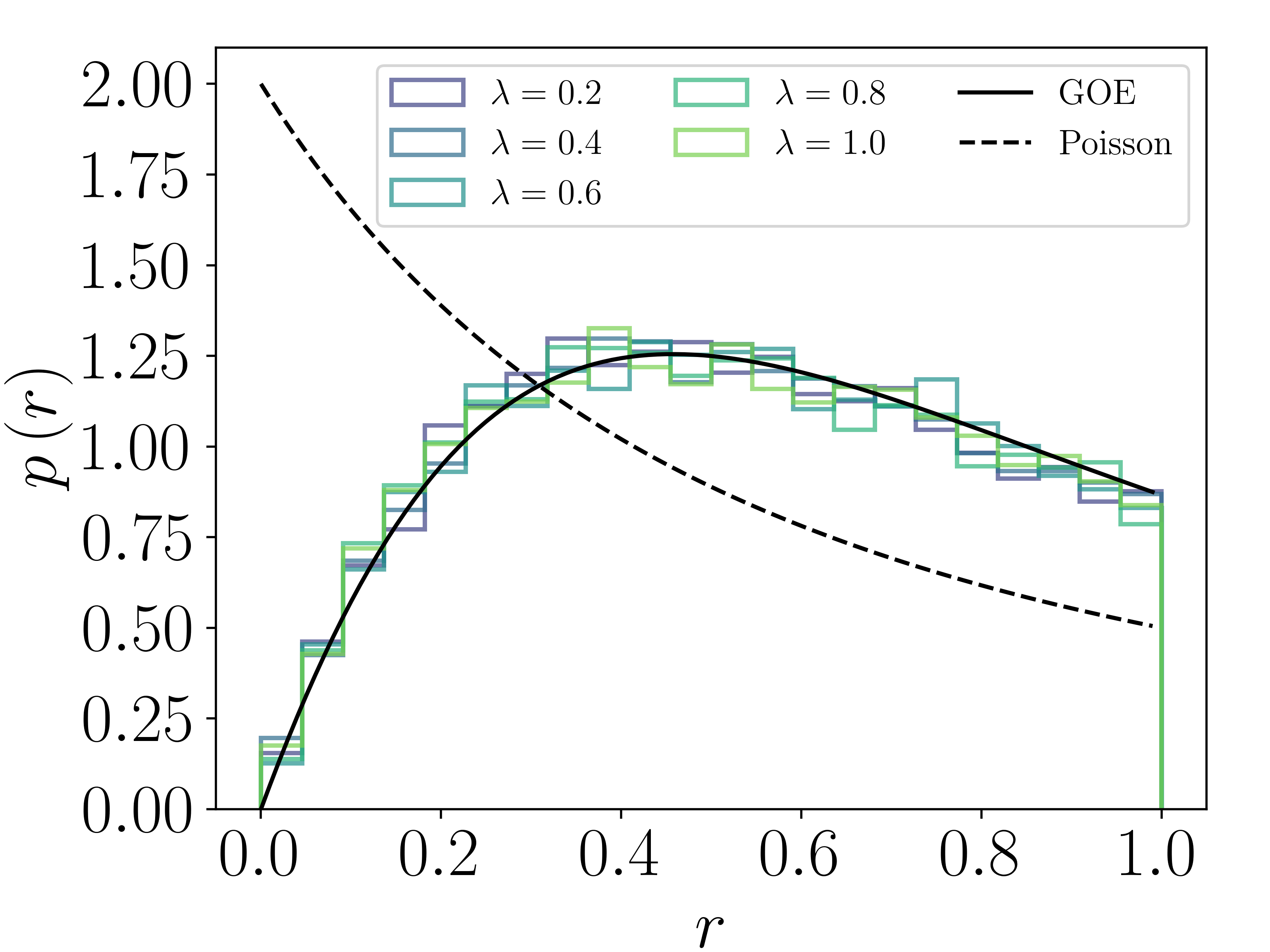}
    \caption{Level spacing distribution for a $12\times 1$ ladder with the height potential at varying $\lambda$. Left: +1 reflection symmetry sector. Right: -1 reflection symmetry sector.     
    Both match the GOE distribution well, indicating that this system is not integrable.}
    \label{fig:level_spacing}
\end{figure}

We start by resolving the symmetries of the system. Translation symmetry is broken due to the open boundaries. The potential breaks charge conjugation symmetry. Reflection across the horizontal axis, as defined above, is identical to charge conjugation and is therefore also broken by the potential. 
Reflection symmetry with respect to the vertical axis needs to be resolved. This results in two sectors, labeled with the eigenvalue $\pm 1$. The distribution $p\left(r\right)$ of consecutive level spacing ratios
\begin{equation}
    r_n = \min\left\{\frac{E_{n+1} - E_n}{E_{n} - E_{n-1}}, \frac{E_{n} - E_{n-1}}{E_{n+1} - E_{n}} \right\}
\end{equation}
of both sectors match the Gaussian Orthogonal Ensemble (GOE)
\begin{equation}
    p_{GOE}(r) = \frac{27}{4} \frac{r + r^2}{(1 + r + r^2)^{5/2}}
\end{equation}
as seen in Fig.~\ref{fig:level_spacing}. This is in contrast to the Poisson distribution expected for integrable models
\begin{equation}
    p_{P}(r) = \frac{2}{(1 + r)^2},
\end{equation}
which is also plotted for reference.

\section{Zero-mode scars}
We refer to zero-modes as eigenstates of the plaquette term of the Hamiltonian with zero eigenvalues. In our notation, these states satisfy
\begin{equation}
    \label{eq:zmodes}
    K \ket{\psi_z} = \sum_n (H^+_n + H^-_n) \left|\psi_z\right> = 0.
\end{equation}
Zero-modes can play a crucial role in forming quantum many-body scars. Systems with a spectral symmetry, together with point-group symmetries, can exhibit an exponential number of zero-modes \cite{schecter2018many}. Their existence follows from an index theorem, which we review and adapt to our models of interest in the Supplementary Material \cite{supplementary}. While one might expect these eigenstates to be thermal, it has been shown that states with low entanglement can be constructed within this subspace for specific cases. It was further conjectured that this is a generic property of local Hamiltonians with an exponential number of zero-modes \cite{karle2021area}. This property was observed explicitly in the case of gauge theories in the spin-$1/2$ QLM, even if it is possible to find scars that are not constructed exclusively from zero-modes \cite{biswas2022scars,sau2023sublattice}.  

We show that it is also possible to construct such states for arbitrary spin truncated link models. The number of these low entropy states grows exponentially with the volume. By choosing the potential term of the Hamiltonian appropriately, different linear combinations can be isolated as low entropy eigenstates.

\subsection{QMBS in TLM With Arbitrary Integer Spin}
\begin{figure}
    \centering
    \includegraphics[width=.35\linewidth]{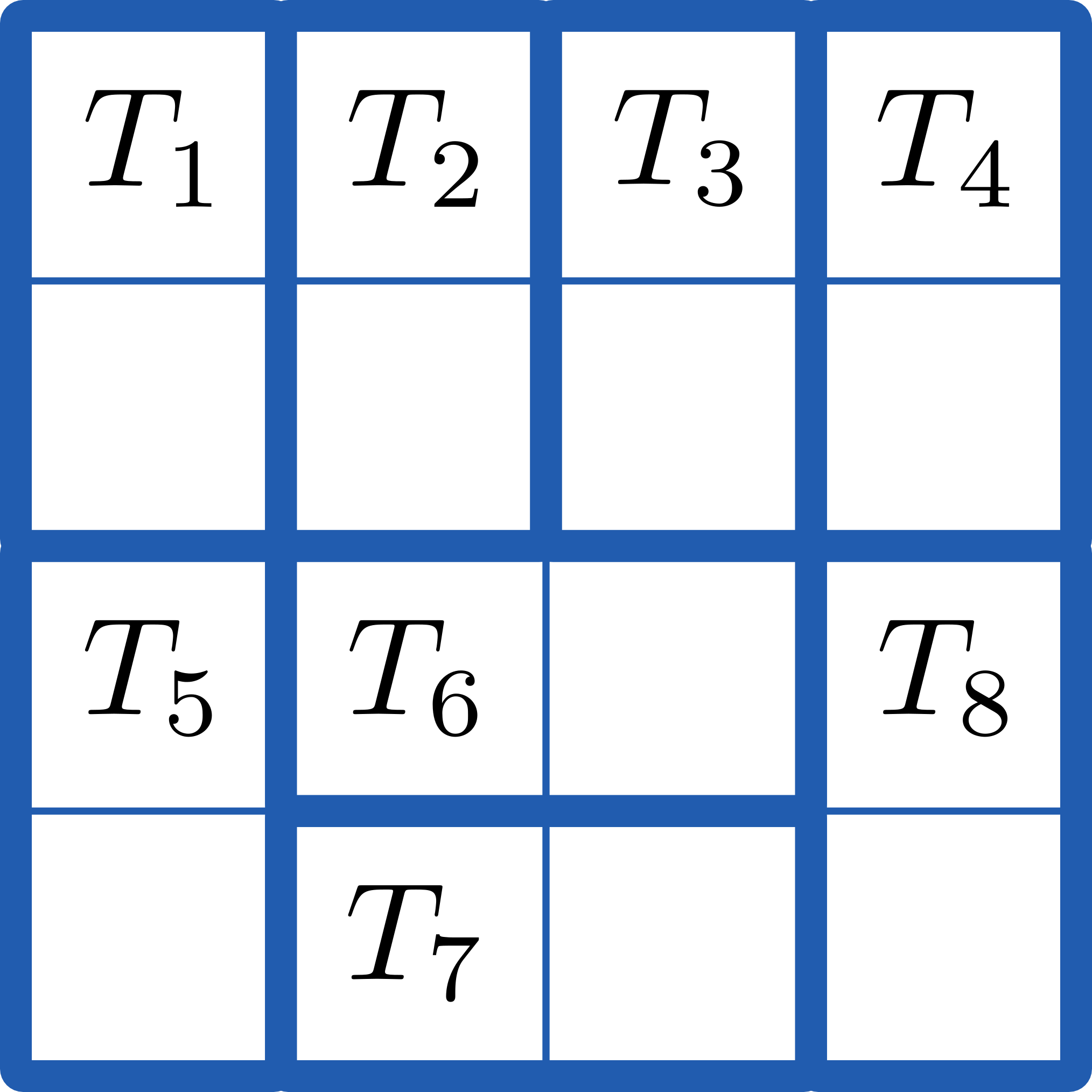}
    \hspace{.1\linewidth}
    \includegraphics[width=.35\linewidth]{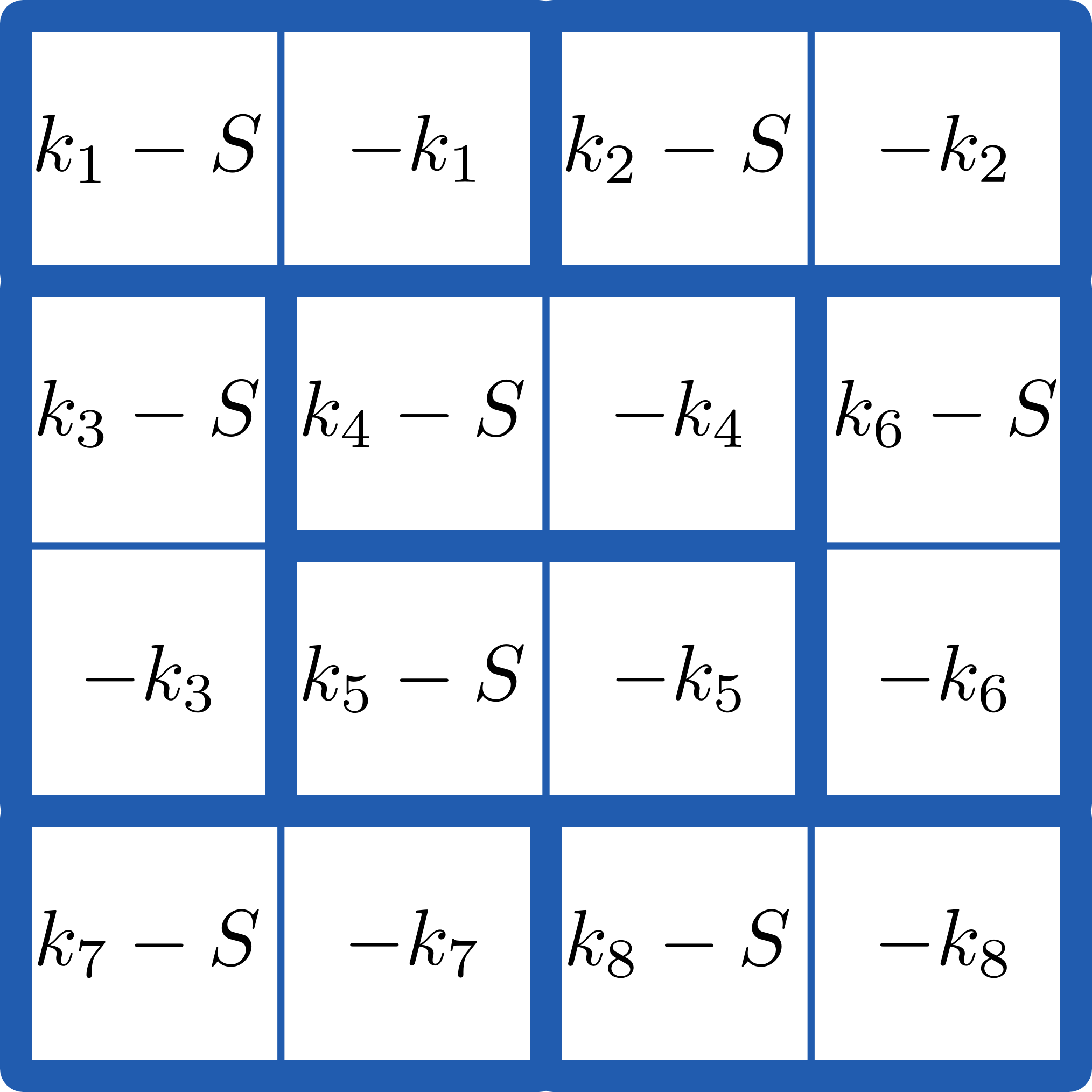}
    \caption{Left: Example of a tiling $T$ of the 2D lattice, consisting of partitioning it into $2\times 1$ and $1\times 2$ tiles, to a total of 8 tiles. Right: Representation of a scar state based on an alternative tiling and in the dual representation. An independent sum over every $k_i$ with a weight of $\prod_i \left(-1\right)^{k_i}$ is assumed. See Eq. \eqref{eq:full-scar-states}.
    }
    \label{fig:tiling}
\end{figure}

Next, we show that zero-modes with area-law entanglement entropy exist in truncated link models of all integer spins. We start by partitioning the 2D lattice into $2\times 1$ and $1\times 2$ tiles, as depicted on the left of Fig.~\ref{fig:tiling}. A given partition will be called tiling and will be denoted by $T$. It can be represented by a set of tuples $\left(n,n'\right)$ where the two entries indicate the two plaquettes making up each tile. We then define the states
\begin{align}
\label{eq:full-scar-states}
    \ket{\psi_s^{(i, T)}} = &\frac{1}{\left(S+1\right)^{\left|T\right|/2}}\prod_{(n,n') \in T} \nonumber \\
    & \left(\sum_{k = 0}^S (-1)^k (H^+_n)^{i-S+k}(H^+_{n'})^{i-k} \right) \ket{\mathbf{0}},
\end{align}
where $\ket{\mathbf{0}}$ is the state where all links have value zero, and $0\leq i \leq S$.
The states $\ket{\psi_s^{(i, T)}}$ have the property 
\begin{equation}
    (H^-_{m} + H^+_{m'}) \ket{\psi_s^{(i, T)}} = (H^+_{m} + H^-_{m'}) \ket{\psi_s^{(i, T)}} = 0
    \label{eq:zm_tile}
\end{equation}
if $(m,m')$ is a tile in $T$. Therefore, 
\begin{equation}
    K \ket{\psi_s^{(i, T)}} = \sum_{(n,n') \in T} (H^+_{n} + H^-_{n'} + H^-_{n} + H^+_{n'}) \ket{\psi_s^{(i, T)}} = 0, 
\end{equation}
making them zero-modes for truncated link models of arbitrary spin. For quantum link models with spin $S>1$, this would no longer be the case due to the electric field-dependent pre-factors when raising/lowering the spins.

These states are zero-modes in all lattices where $L_1 L_2$ is even. The number of possible tilings grows exponentially with $L_1 L_2$ if both $L_1$ and $ L_2$ are larger than 1.

By partitioning the systems into two regions, we see that the entanglement between them is generated by tiles that touch both regions resulting in an area law. 
The states $\ket{\psi_s^{(i, T)}}$ are then mid-spectrum states with area law entanglement entropy, making them quantum scars.

To get a deeper understanding of the structure of these scars, and to connect them to lego scars that have been constructed in the $S=1/2$ QLM, it is useful to define the \emph{dual basis} for integer spin.
Height variables $h_n  \in \mathbb{Z}$ live at the center of plaquettes. The values of the vertical links are given by the difference of the height variables of the plaquettes to their right and left, while the values of the horizontal links are given by the difference of the height variables of the plaquettes to their bottom and top, i.e. $E_{n1} = h_{n-\hat{2}} - h_n$ and $E_{n2} = h_n - h_{n-\hat{1}}$. Height variable configurations with $|h_n - h_{n'}| \leq S$ for all neighboring plaquettes $n$, $n'$ represent a valid spin-$S$ link configuration.

The height variable representation with periodic boundary conditions is not unique, since adding a constant to all height variables gives the same state.
If there is an open boundary condition in at least one direction, we set all height variables at one open boundary to have the value of the neighboring boundary link. In our construction, we always choose the top boundary, which makes the representation unique.
The kinetic operators $H^+_n$ and $H^-_n$, are raising/lowering operators for the height variable $h_n$. 

\begin{figure}
    \centering
    \includegraphics[width=.7\linewidth]{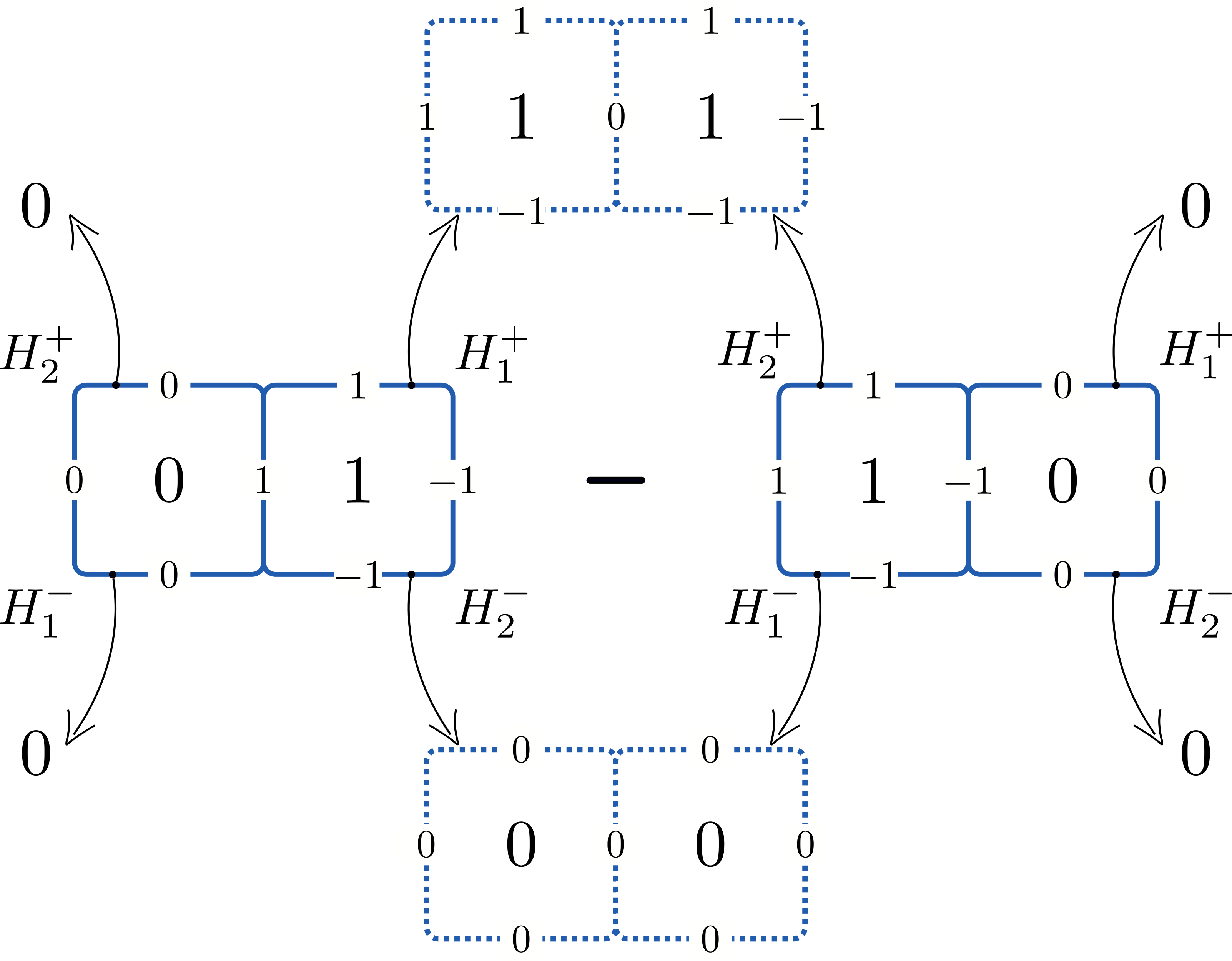}
    \caption{Schematic depiction of the scar building block for $S=1$. The tiles depicted with solid lines make up the scar state. The operators $H^\pm_n$ raise or lower the height variables in plaquette $n$. They either annihilate the state or generate another state, depicted with dashed lines. This is canceled by applying the other type of operator on the second state.}
    \label{fig:two_pq}
\end{figure}

In the dual representation, the state  \eqref{eq:full-scar-states} in a $2\times 1$ plaquette system is
\begin{equation}
    \label{eq:scar_tiles}
    \ket{\psi_i} = \frac{1}{\sqrt{S + 1}} \sum_{k = 0}^{S} (-1)^k \ket{(i-S+k)\ (i-k)}
\end{equation}
where the state is labeled with its height variables $\ket{h_1\ h_2} = (H^+_1)^{h_1} (H^+_2)^{h_2} \ket{\mathbf{0}}$.
An illustration of this tile and the terms that cancel each other in the sum $K\ket{\psi_i}$ for $S=1$ can be found in Fig.~\ref{fig:two_pq}.

All states that contribute in \eqref{eq:full-scar-states} have a height variable representation where there is one state out of the sum \eqref{eq:scar_tiles} on all tiles of the tiling $T$. The state sums over all $(S+1)^{\left|T\right|}$ combinations of these states.
A pictorial representation of this tiling is shown on the right-hand side of Fig.~\ref{fig:tiling}.
This tiling structure is similar to the lego and sub-lattice scars constructed for the $S=1/2$ case \cite{biswas2022scars, sau2023sublattice}, and allows for the construction of an exponential number of scars for arbitrary truncated link models. A formal definition of this tiling, in the form of a tiling product, can be found in the Supplementary Material \cite{supplementary}. There, we also describe a more general framework to construct more low-entropy zero-modes in TLMs. 

\subsection{Beyond Zero-Mode Building Blocks}
\label{sec:beyondZMtiles}

The above-described mechanism follows two basic ingredients. It constructs zero-modes from a basic tile made of two plaquettes joined together in one link. The full state is obtained by tiling these building blocks together. In this section, we demonstrate that it is possible to form zero-modes with different structures.
This demonstrates that even though we have constructed numerous scars for arbitrary spin, there are additional routes to obtain other types of zero-mode scars. We provide explicit constructions for the $S=1$ case. The two examples described here exhibit a sublattice structure similar to what has been found for spin-$1/2$ \cite{biswas2022scars,sau2023sublattice}. 
One can be constructed by making use of diagonal tiles, while the other does not exhibit any tiling structure. A third example, that we have found for the specific case of a $4\times 4$ volume, can be found in the Supplemental Material \cite{supplementary}. To the best of our knowledge, that specific construction is not generalizable for arbitrary volumes.

\paragraph{Diagonal Tiling}

\begin{figure}
    \centering
    \raisebox{-0.5\height}{
    \includegraphics[width=0.395\linewidth]{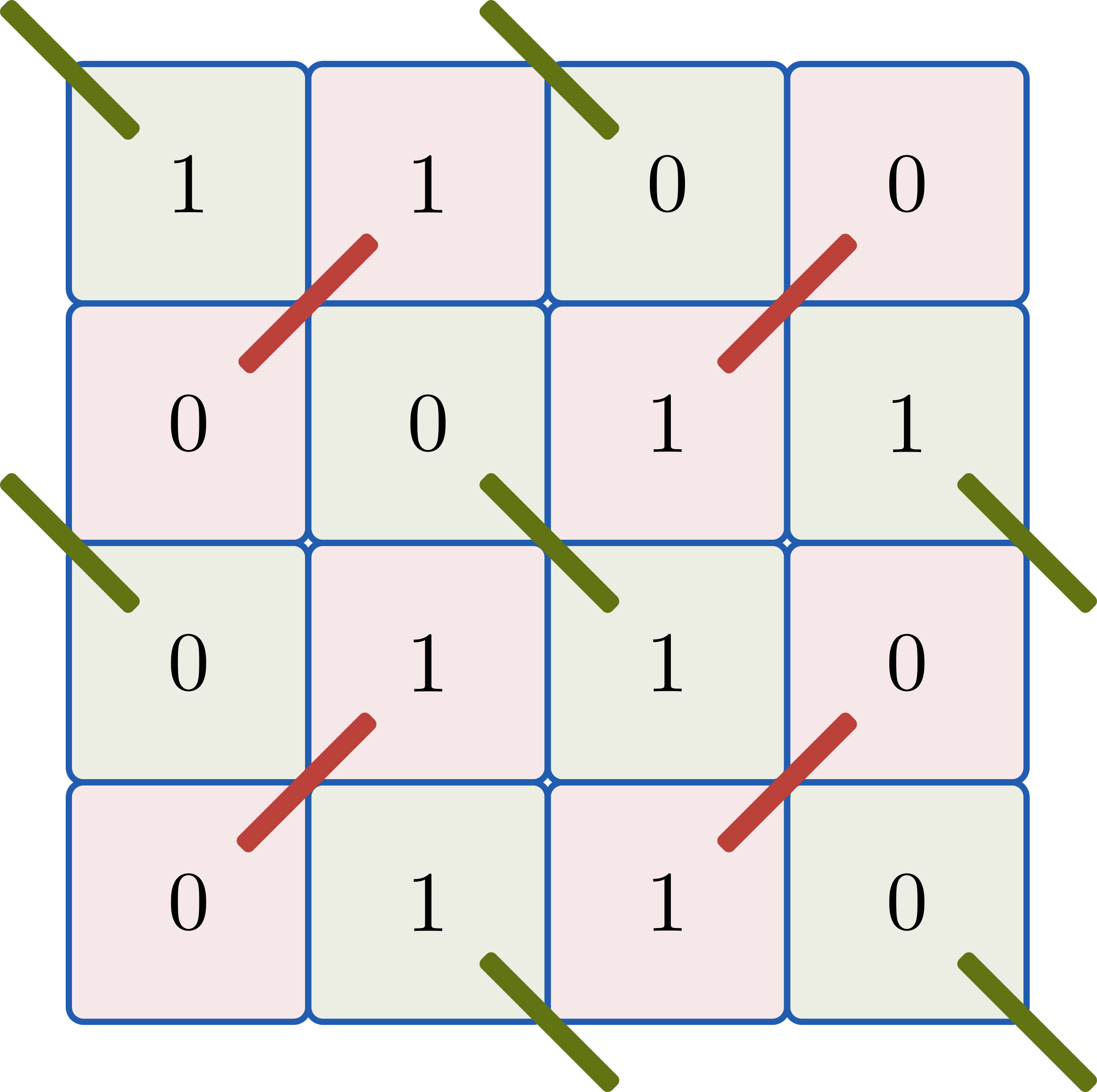}}
    \hspace{.05\linewidth}
    \raisebox{-0.5\height}{\includegraphics[width=0.35\linewidth]{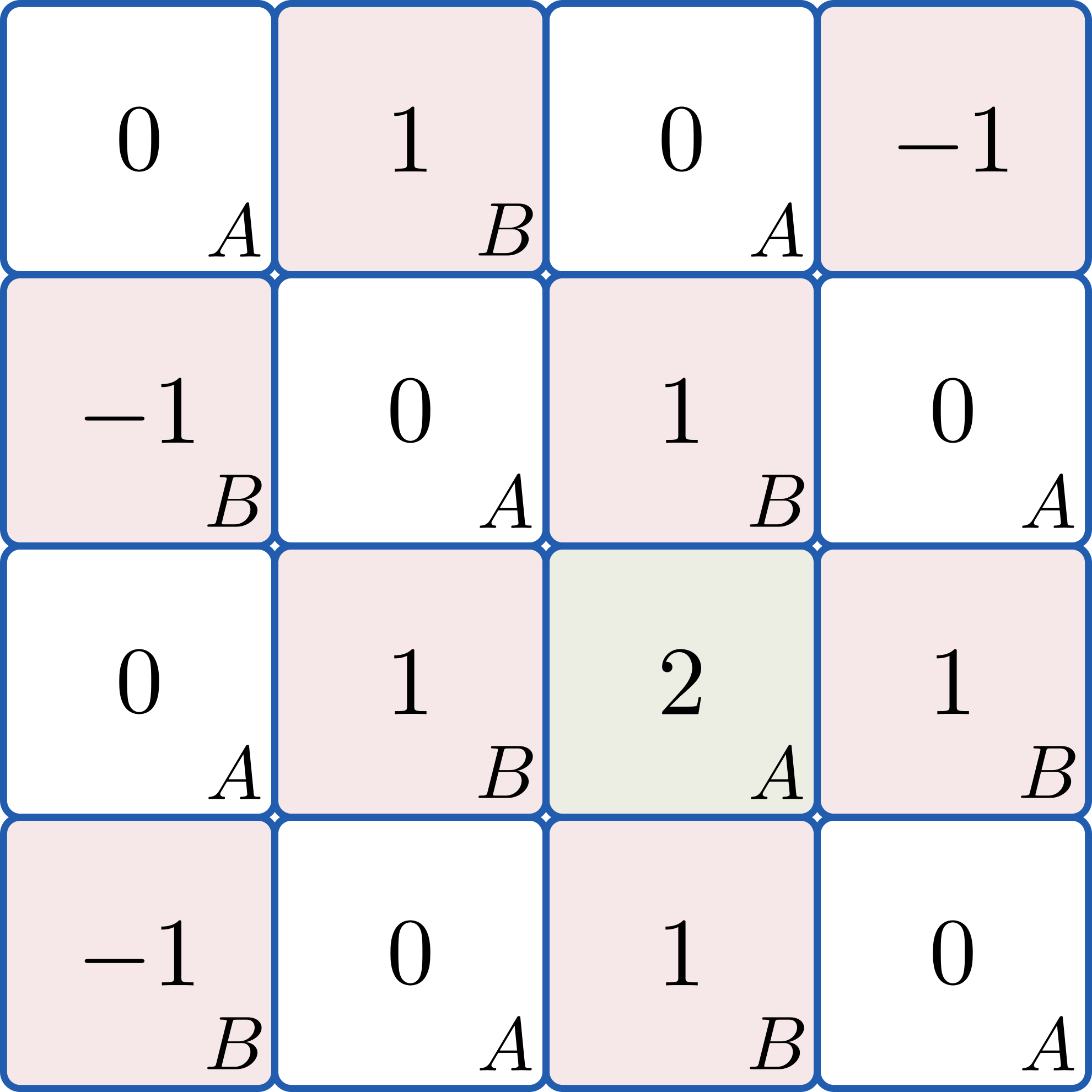}}

    \caption{Illustrations of $S=1$ zero-mode scars that are eigenstates of $\sum_n E_n^2$, exhibiting sublattice structure. Left: Scar built by singlet states of two plaquettes connected diagonally. Plaquettes connected by red or green lines form tiles.
    Right: Singlet states are placed in sublattice $B$ while sublattice $A$ is filled by 0's. If all plaquettes surrounding a 0 have the same value of $\pm1$, then that plaquette is combined with $\pm2$.}
    \label{fig:sublattice_scar}
\end{figure}

We consider a scar with the generic form previously described in \eqref{eq:full-scar-states} for the specific case of spin $S=1$, i.e.
\begin{align}
    \ket{\psi_s^{(\pm, T)}} = &\frac{1}{2^{\left|T\right|/2}}\prod_{(n,n') \in T}\left(\ket{\pm1\ 0} - \ket{0\ \pm1}\right),
    \label{eq:diagonal_tiling}
\end{align}
in the dual representation. The two plaquettes that make up tiles no longer share a link, but neighbor each other diagonally in a pattern depicted in Fig.~\ref{fig:sublattice_scar}. Individual tiles are not zero-modes, i.e. they do not satisfy \eqref{eq:zm_tile}. However, the resulting state is a zero-mode of the full system. This follows because each plaquette neighbors both sites of another tile. One of the plaquettes of this tile will take the value $\pm 1$. None of the plaquettes can therefore ever take height $\mp 1$. 

\paragraph{Non-Tiling Scar}

It is also possible to have scars that do not follow \eqref{eq:full-scar-states} nor exhibit a similar tiling structure. We divide the lattice into two sublattices $A$ and $B$ and construct a zero-mode according to

\begin{align}
    \ket{\psi_s}=\frac{1}{2^{L_1L_2/2}}
    &\prod_{m \in A}\left(1-\left(H^+_m\right)^2P^+_m-\left(H^-_m\right)^2P^-_m\right) \nonumber \\
    &\prod_{n\in B}\left(H^+_n-H^-_n\right)\ket{\bold{0}},
    \label{eq:non_tiling_scar}
\end{align}
where $P^\pm_m$ is the projector on $\pm 1$ on all plaquettes surrounding the plaquette at $m$. An illustration can be found in Fig. \ref{fig:sublattice_scar}.
This state can be understood by considering one sublattice and creating a linear superposition of heights $\pm1$. In our example, this is done for sublattice $B$. For plaquettes of sublattice $A$, the height variable is stuck at zero as long as all surrounding plaquettes are not equal. Otherwise, the plaquette can either be raised or lowered depending on the value of the surrounding plaquettes being positive or negative. By combining those states with the respective lowered and raised states with the opposite sign, we obtain a zero-mode.

\section{Numerical results and discussion}
While the derived scars are low entropy zero-modes, they are degenerate with all the other zero-modes and are not easily identifiable in the spectrum. In the following, we introduce potentials that have these states as eigenstates. By analyzing the Shannon entropy and the bipartite entanglement entropy of the eigenstates of the Hamiltonian, we verify the existence of these low entropy states numerically. Concretely, we calculate the Shannon entropy with respect to the electric field basis $\{\ket{\phi_i}\}$. For a state $\ket{\psi} = \sum_i c_i \ket{\phi_i}$, the Shannon entropy has the value
\begin{equation}
    S = -\sum_i |c_i|^2 \log |c_i|^2.
\end{equation}
For the entanglement entropy, we split the system into two subsystems $A$ and $B$ of equal size. The bipartite entanglement entropy is then given by,
\begin{equation}
    S_{\frac{L}{2}} = -\text{Tr} \rho_A \log \rho_A = -\text{Tr} \rho_B \log \rho_B.
\end{equation}
Explicitly, we always partition the system by dividing it vertically into two halves. We consider the bordering vertical links on the left as belonging to the subsystem while the ones on the right do not. See e.g. the Supplementary Material of \cite{banerjee2021quantum} for details on how to calculate the entanglement entropy in these systems.

We start by isolating the scars of the form of~\eqref{eq:full-scar-states}. We observe that the sum of the height variables remains constant and equal to $(2i-S)$ for each term. This does not mean that the sum over height variables is a valid potential because, in general, the height variables are not uniquely defined from the physical configuration. However, they can be made unique for open boundary conditions, as described above. For example, in the case of a single-leg ladder, the sum of all height variables can be seen as the sum of the horizontal links of the top row. In general, for open boundary conditions, we write
\begin{align}
    V = \lambda \sum_n h_n,
    \label{eq:Ladder-potential}
\end{align}
which will have the constructed scars as eigenstates with energy $E = \lambda \frac{L_1 L_2}{2} (2 i - S)$.

\begin{figure}
    \centering
    \includegraphics[width=0.49\linewidth]{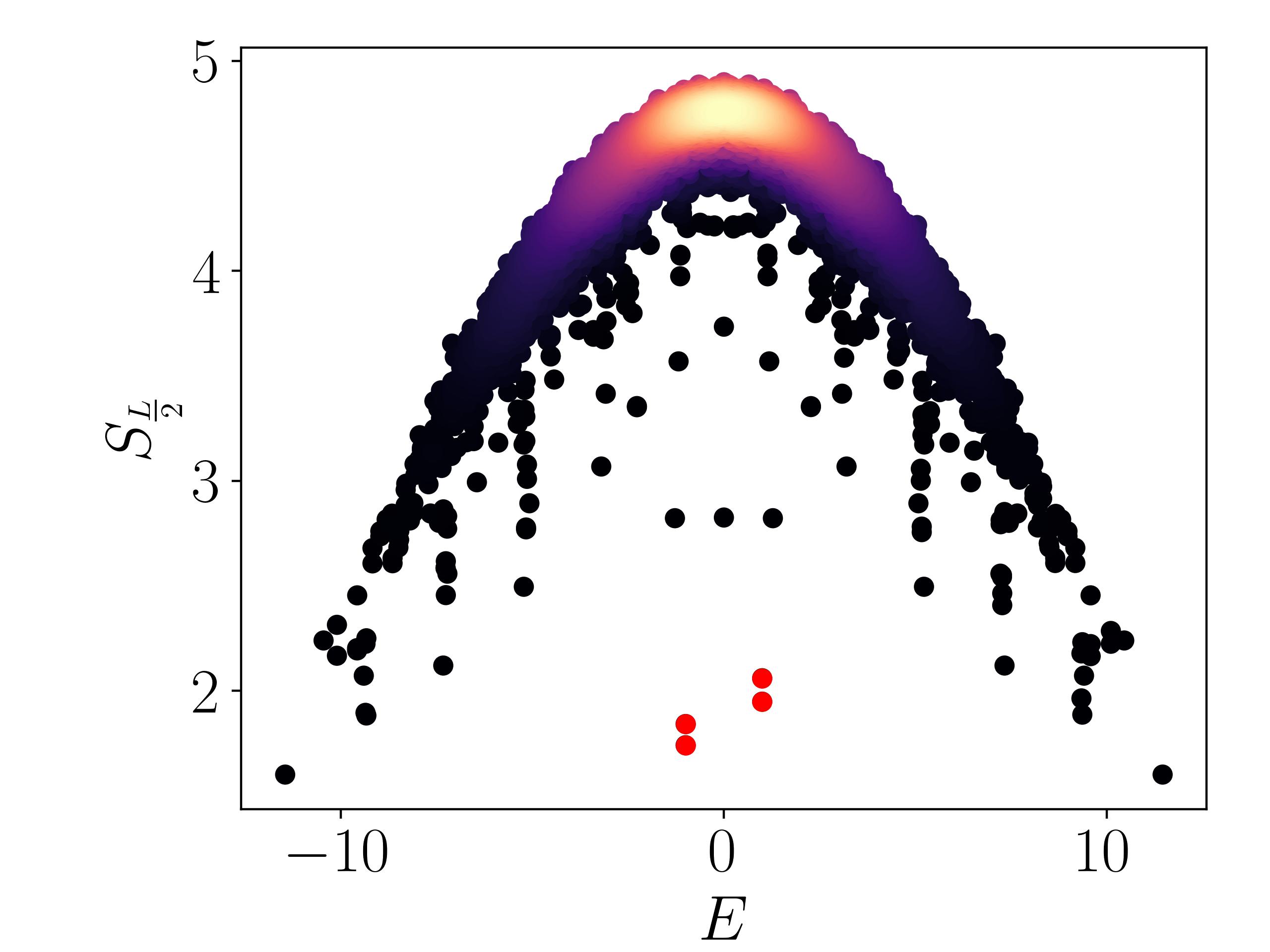}
    \includegraphics[width=0.49\linewidth]{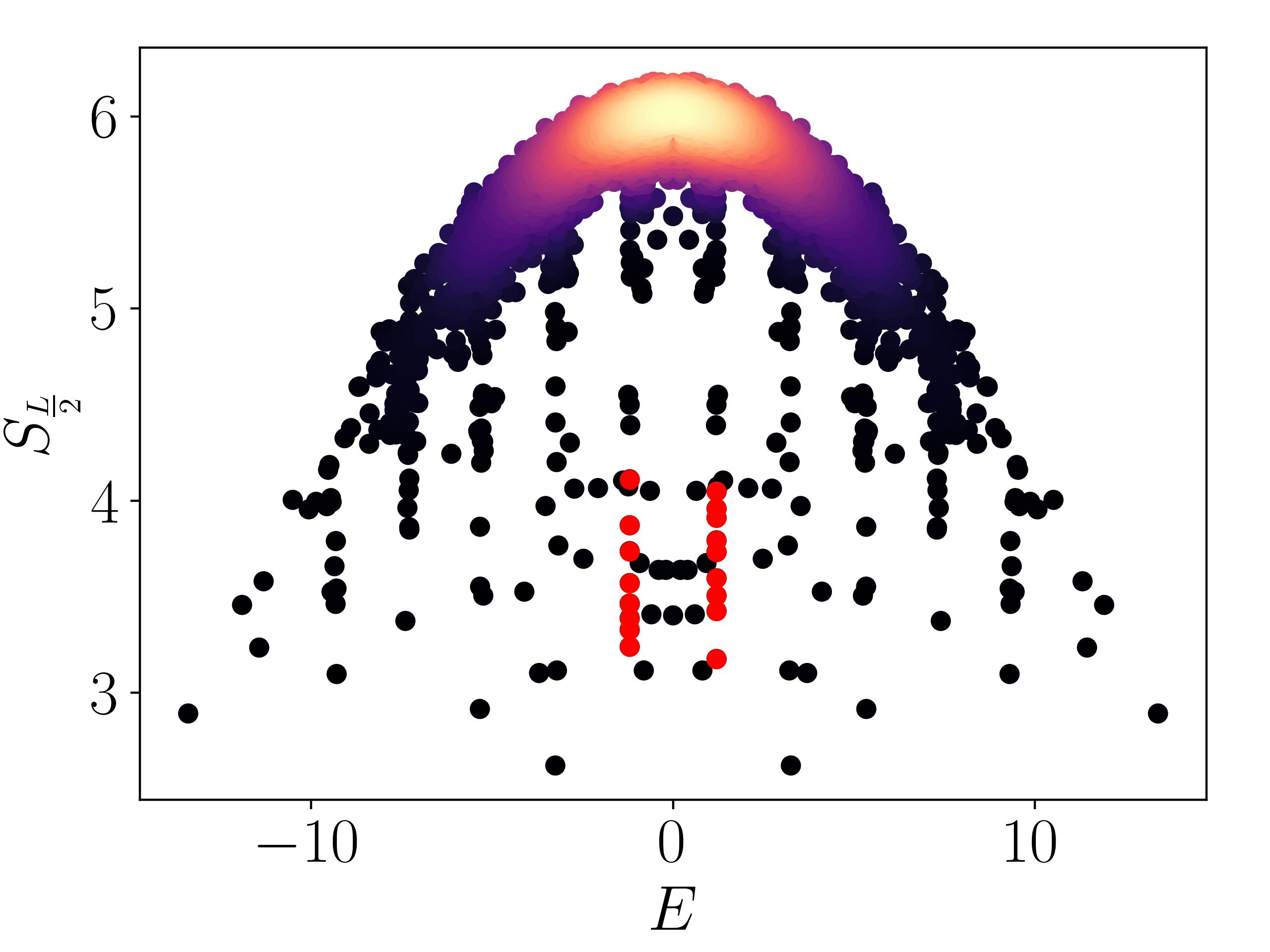}
    \caption{
    Entanglement entropy for two different volumes with periodic boundary conditions in the $x$ direction, $S=1$, and height potential \eqref{eq:Ladder-potential} with $\lambda=0.2$. Brighter colors indicate a higher density of states.
    Left: $10 \times 1$ ladder. Four mid-spectrum states with low entropy, shown in red, are visible. They are superpositions of the predicted states.
    Right: $6 \times 2$ volume in the zero momentum sector. Many mid-spectrum states with low entropy are present. The towers marked in red contain the predicted states.
    }
    \label{fig:s1ladder}
\end{figure}

For $S=1$ the two zero-mode tiles are
\begin{align}
    \ket{\psi_\pm} = \frac{1}{\sqrt{2}}(\ket{0\ \pm1} - \ket{\pm1\ 0}),
    \label{eq:spin1tile}
\end{align}
which correspond to choosing $i=1$ and $i=0$ in \eqref{eq:scar_tiles}.
We first study a $S=1$ ladder with $L\times 1$ plaquettes. This is a chain of plaquettes with periodic boundary conditions in the direction of the chain, but not the direction perpendicular to it.  For such a ladder, there are exactly two distinct tilings of $2\times 1$ tiles. They correspond to placing the lower left corner of the tiles in either even or odd sites. These two states, coming from two different tilings with the same $i$, are not orthogonal but generate a two-dimensional scarred Hilbert space.  
We then expect four scars. Two degenerate low entropy states with energy $L\lambda/2$ from the tilings of state $\ket{\psi_+}$ ($i=1$) and two with energy $- L\lambda/2$ from tilings of state $\ket{\psi_-}$ ($i=0$).
This is exactly what is observed when performing exact diagonalization on small systems, the results of which are presented in Fig.~\ref{fig:s1ladder}. 
The four low-entropy states are clearly identifiable and each of them is composed of the expected states, as shown in the Supplementary Material \cite{supplementary} by looking at the amplitude of the scars in the electric field basis.

For systems with a larger vertical direction, there are many more ways to tile the plane with $2 \times 1$ and $1\times 2$ tiles. Adding a second leg is enough to make the number of possible tilings exponentially large, all of which will mix in the spectrum. 
Using the sum of the height variables as the potential, we expect towers of low-entropy states at energies $\pm L_1L_2 \lambda/2$.
These two towers are observed, as depicted in Fig.~\ref{fig:s1ladder}. There are many other low-entropy states at other energies. While we observed these states to be built from zero-mode tiles, we do not expect their construction to be generalizable beyond $2\times L$ systems and will not attempt a detailed characterization here. 

\begin{figure}
    \centering
    \includegraphics[width=0.49\linewidth]{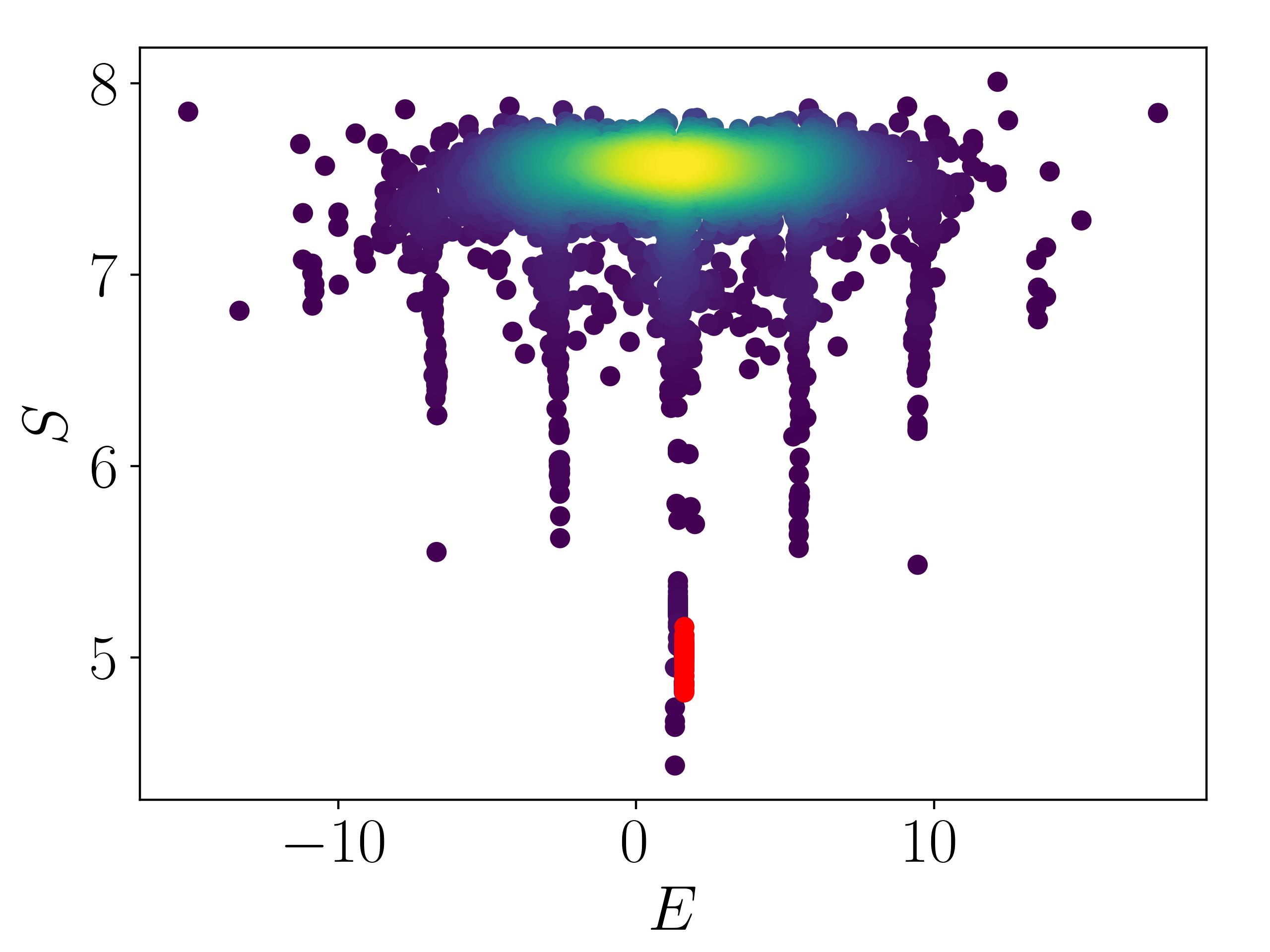}
    \includegraphics[width=0.49\linewidth]{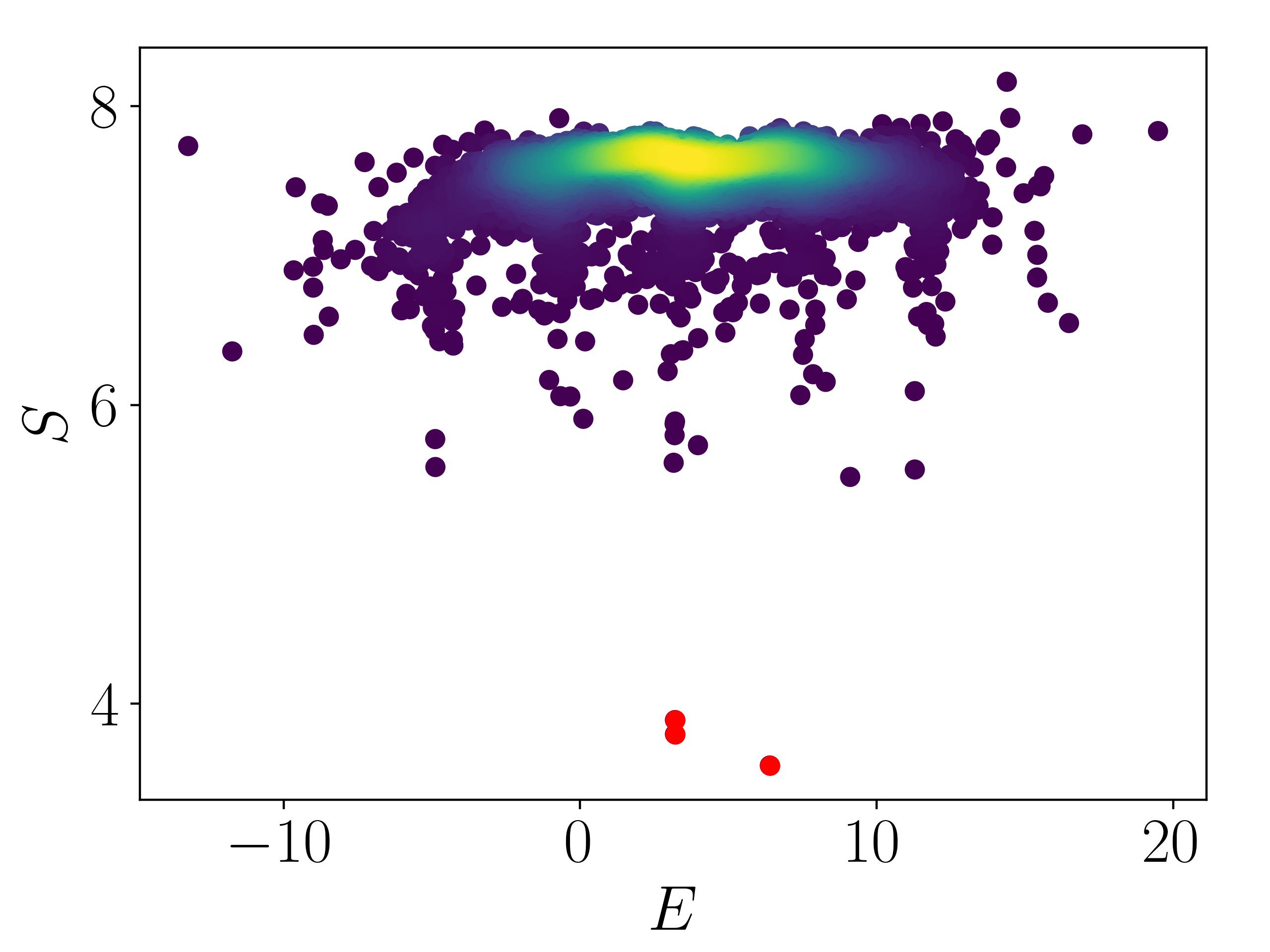}
    \caption{
    Shannon entropy for a $4\times 4$ system with periodic boundary conditions in both directions, $S=1$, and different potentials. Shown states are in the zero momentum and $+1$ charge conjugation sectors. Bright colors indicate a higher density of states.
    Left: Hamming distance potential with $\lambda = 0.2$.  Many mid-spectrum states with low entropy are visible. The tower colored in red at $E=1.6$ contains the predicted scars.
    Right: Standard electric field term \eqref{eq:E2-potential} with $g = 0.2$. Three mid-spectrum states with low entropy are visible and colored in red.}
    \label{fig:E2-scar}
\end{figure}

All scars we have constructed also exist in periodic boundary conditions. Using the sum of the height variables as the potential is not well defined, as previously discussed.
Instead, for the scars of the form \eqref{eq:full-scar-states}, we use the minimum number of kinetic operators that must be applied to the state $\ket{\textbf{0}}$ to arrive at any given electric field basis state $\ket{\phi_i}$. This is similar to the Hamming distance as used in e.g. \cite{desaules_hypergrid_2022} generalized for higher spin. We therefore call this the Hamming distance potential.
The entropy using this potential is shown in the left panel of Fig.~\ref{fig:E2-scar}, where the tower of predicted scars is visible.
Additional towers are present at non-integer multiples of $\lambda$, which are not characterized here.

For the other types of scars described for $S=1$, all contributing states in~\eqref{eq:diagonal_tiling} and~\eqref{eq:non_tiling_scar} have the same number of links with electric field zero. In the first case, half of the links are zero while in the second no link is zero. These scars can then be isolated by the standard electric field term
\begin{equation}
    \label{eq:E2-potential}
    V = g \sum_n E_n^2.
\end{equation}
The third scar, described in the Supplemental Material \cite{supplementary}, is also an eigenstate of this potential. These are precisely the three scars observed in the right panel of Fig.~\ref{fig:E2-scar}.

\begin{figure}
    \centering
    \includegraphics[width=0.49\linewidth]{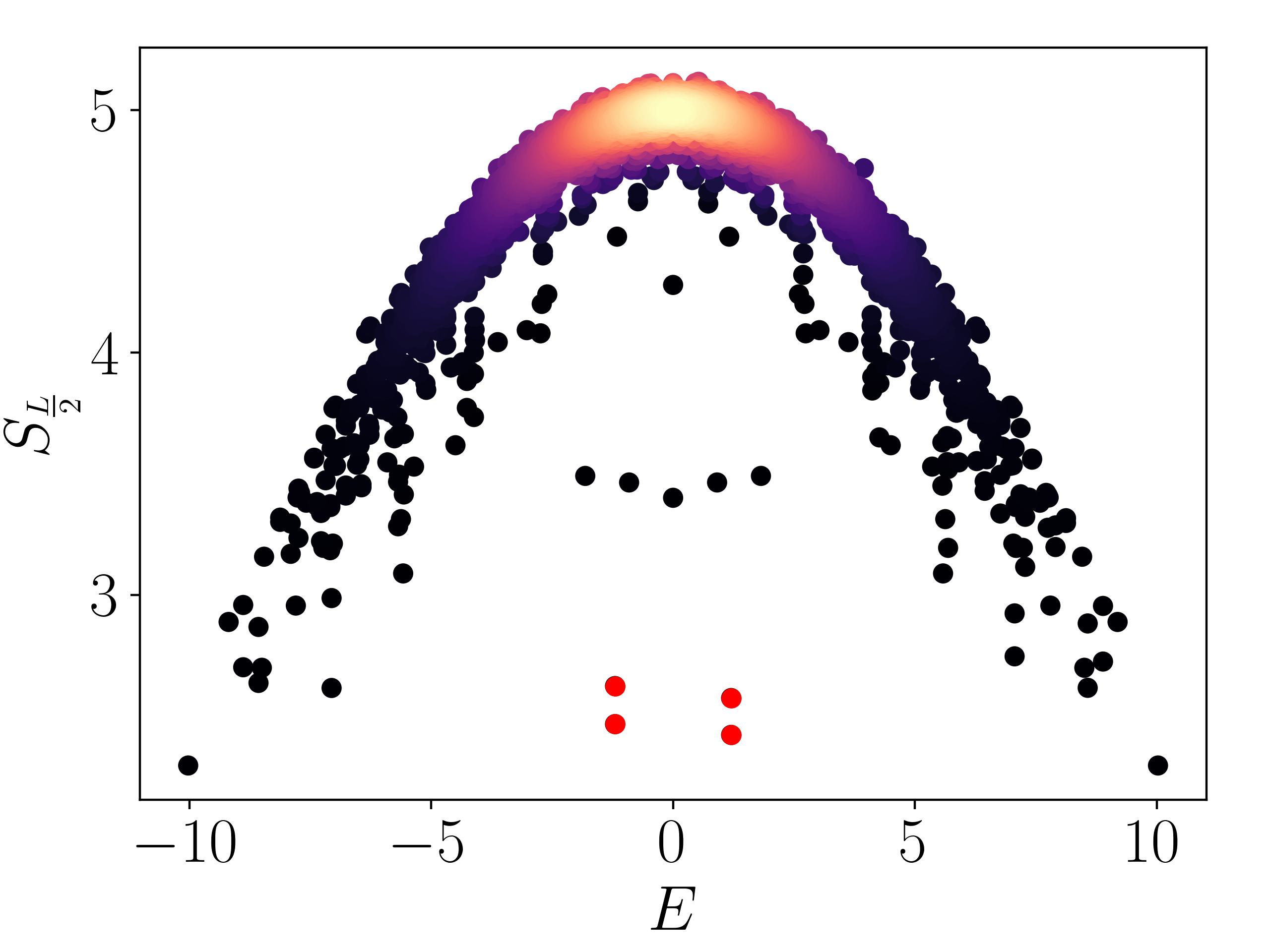}
    \includegraphics[width=0.49\linewidth]{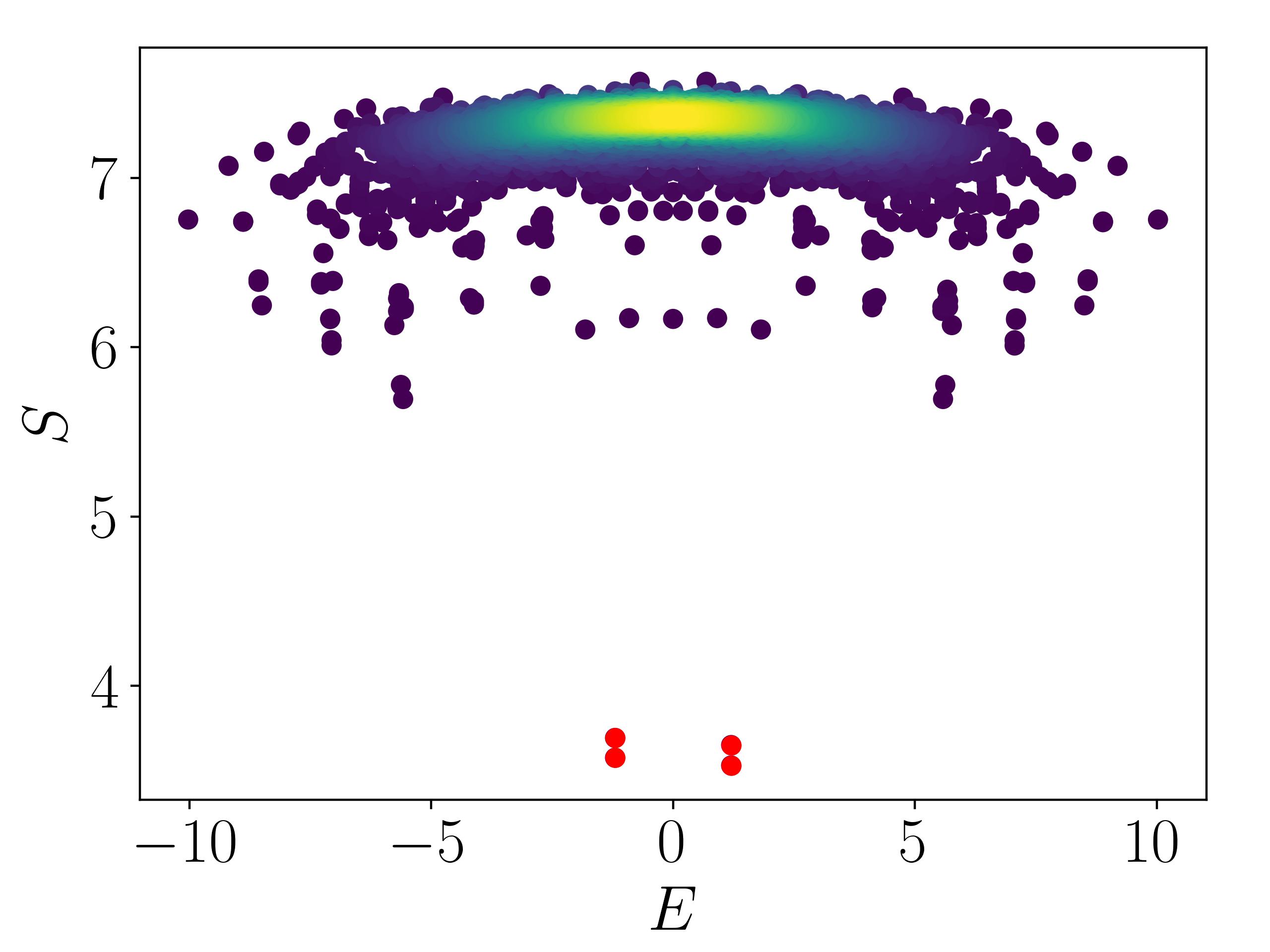}
    \caption{
    Entropies for a $S=2$ ladder with 6 plaquettes and $\lambda = 0.2$. Brighter colors indicate a higher density of states.
    Left: Entanglement entropy.  Right: Shannon entropy.  Four mid-spectrum states with low entropy are visible, colored in red. The remaining two predicted states are mixed with a tower of zero-modes at $E=0$.
    }
    \label{fig:s2ladder}
\end{figure}

For $S=2$ we have studied the $6 \times 1$ ladder in Fig.~\ref{fig:s2ladder} with the height potential~\eqref{eq:Ladder-potential}. In this case, we expect scars obtained from choosing $i=0$, $i=1$, and $i=2$ in \eqref{eq:scar_tiles}. The $i=1$ scar has zero energy for any value of $\lambda$, while the other two will have $E_{2,1} = \pm 6\lambda$. The latter two are evident in the figure. The $i=1$ scar is not visible, since it is in superposition with a tower of zero-modes. 

This numerically demonstrates the existence of low entanglement entropy zero-modes in systems beyond $S=1$. 
For larger volume or higher spin systems, we have established the existence of such states analytically in the earlier sections of this article.

\section{Conclusions and outlook}
In this work, we demonstrate that Quantum Many-Body Scars (QMBS) exist across numerous 2D Abelian pure gauge theories. In the limit of zero coupling, a spectral symmetry emerges, leading to an exponentially large number of zero-modes. We show that it is possible to construct many states with area-law entanglement entropy within this subspace. Since these are mid-spectrum states of a non-integrable model, they constitute QMBS. The number of scars we could identify grows exponentially with the volume, but comprise an exponentially small fraction of the Hilbert space. This indicates weak breaking of the Eigenstate Thermalization Hypothesis. 

Through the analytical construction of the scars, we can strategically add a potential to the Hamiltonian that isolates them from other eigenstates. This allows for the numerical verification of their presence for moderate volumes and values of the spin. Different choices of the potential can isolate different scars. If the new potential does not have scar states as eigenstates, they will disappear from the spectrum. An interesting question for future work is whether this type of analytical construction can be generalized in the presence of bosonic or fermionic matter.

For $S=1$, some scars survive the presence of the standard $g\sum_nE^2_n$ potential, if the lattice size is even in both directions and under periodic boundary conditions. It would be interesting to understand if a similar dependence on the geometry could occur in other gauge theories, such as $SU\left(2\right)$, where scars were not found for larger truncations in single-leg ladders \cite{ebner2024entanglement}.

Our analytical construction of QMBS extends to arbitrary truncated spins and arbitrarily large volumes.  This result establishes the presence of scars well beyond the reach of existing numerical methods. 
It can be used to guide future quantum simulations of gauge theories, in a regime beyond classical simulations, towards interesting non-equilibrium phenomena. From the theoretical point of view, it constitutes a step towards understanding the role of scars when taking the continuum limit, where the Hilbert space per link is not bounded.

{\it Acknowledgements.}---We are grateful to Debasish Banerjee for insightful discussions. T.B. thanks the Galileo Galilei Institute for Theoretical Physics for the hospitality during the completion of part of this work. This research was supported by the Munich Institute for Astro-, Particle and BioPhysics (MIAPbP), which is funded by the Deutsche Forschungsgemeinschaft (DFG, German Research Foundation) under Germany´s Excellence Strategy – EXC-2094 – 390783311. We acknowledge access to Piz Daint at the Swiss National Supercomputing Centre, Switzerland under the ETHZ’s share with the project ID eth8. Support from the Google Research Scholar Award in Quantum Computing and the Quantum Center at ETH Zurich is gratefully acknowledged.

{\it Note.}---During the final stages of our manuscript, we became aware of another work~\cite{osborne2024quantum} on quantum many-body scarring in a 2+1D U(1) gauge theory with dynamical matter.~\\

\bibliography{apssamp}

\setcounter{section}{0}
\begin{center}
    \Large{\textbf{Supplementary Material}}
\end{center}

\section{Symmetries of the systems}

The symmetries of the Hamiltonians considered will depend on the choice of the potential. We will always have 

\begin{itemize}
    \item \textbf{Translation Symmetry} along the horizontal axis $E_{ni}\rightarrow E_{n+\hat{1}i}$, $U_{n+\hat{1}i}\rightarrow U_{n+\hat{1}i}$. Translation along the vertical axis is not always present due to boundary conditions;

    \item \textbf{Reflection Symmetries} with respect to the two axis ${\cal I}_x$ and ${\cal I}_y$. Explicitly, ${\cal I}_x$ is given by $E_{\left(n_1,n_2\right)1}\rightarrow E_{\left(n_1,-n_2\right)1}$, $E_{\left(n_1,n_2\right)2}\rightarrow -E_{\left(n_1,-n_2\right)2}$ $U_{\left(n_1,n_2\right)1}\rightarrow U^\dagger_{\left(n_1,-n_2\right)1}$, $U_{\left(n_1,n_2\right)2}\rightarrow U^\dagger_{\left(n_1,-n_2\right)2}$ and analogous for ${\cal I}_y$. These symmetries are important to prove the existence of an exponential number of zero-modes;

    \item \textbf{Charge Conjugation}, which is characterized by the field transformations $E_{ni}\rightarrow -E_{ni}$ and $U_{ni}\rightarrow U^\dagger_{ni}$. 
\end{itemize}

When explicitly referenced, we restricted our exact diagonalization to the zero momentum sector and to the +1 charge conjugation sector. 
Otherwise, the plots are in the electric field basis and no symmetries beyond applying Gauss' law and being restricted to the zero winding sector are used.

\section{Index Theorem and an Exponential Number of zero-modes}

In \cite{schecter2018many} it was shown that non-integrable models can still exhibit an exponential number of eigenstates with zero energy. We review this result and adapt it to our models of interest. For this section, we will refer to plaquette flipping terms as the kinetic part of the Hamiltonian

\begin{equation}
    K=\sum_{n}\left( U_{n1}^\dagger U_{n+\hat{1}2}^\dagger U_{n2} U_{n+\hat{2}1}+\mathrm{h.c.}\right).
\end{equation}
We only need to assume that $U_{ni}/U_{ni}^\dagger$ are operators that raise/lower the value of the electric field on the link by 1. This makes the result very robust with respect to a wide range of formulations.

We define a set of link operators $\zeta_{ni}$, which are diagonal in the electric field basis. For each link we have

\begin{equation}
    \zeta\ket{\varepsilon}=
    \left(-1\right)^\varepsilon
    \ket{\varepsilon}.
\end{equation}
We further construct

\begin{equation}
    \label{eq:antic}
    {\cal C}=
    \prod_{n=0}^{L_1-1}
    \prod_{a=0}^{L_2/2-1}
    \zeta_{\left(n,2a\right)1},
\end{equation}
where we have assumed $L_1$ to be even.
We can see that ${\cal C}$ anti-commutes with the kinetic Hamiltonian $\left\{{\cal C},K\right\}=0$. This implies that if $\ket{E}$ is an eigenstate with energy $E$, then ${\cal C}\ket{E}$ has energy $-E$. In the subspace generated by eigenstates with zero energy, we can diagonalize ${\cal C}$ and $K$ together. 

Let now ${\cal I}_y$ be the unitary transformation that reflects along the vertical axes, as defined above. This is a symmetry of the Hamiltonian and also commutes with ${\cal C}$. We can now use the existence of ${\cal I}_y$ and ${\cal C}$ to show the existence of an exponential number of zero-modes.

First, note that the number of zero-modes $N_0$, is bounded from below by $N_0\geq\left|\mathrm{tr}\left({\cal CI}_y\right)\right|$. We can see this by adopting a basis that diagonalizes both $K$ and ${\cal I}_y$. Such eigenstates are always mapped to a different energy eigenstate by ${\cal C}$ unless the energy is zero. As a consequence, only zero-modes contribute to the trace. Each eigenstate, on that sector, will contribute with a value in the interval $\left[-1,1\right]$. This shows that the number of zero-modes can never be smaller than $\left|\mathrm{tr}\left({\cal CI}_y\right)\right|$.

We can now show that the trace is actually exponential on the volume by considering the electric field basis. These are all eigenstates of ${\cal C}$, but the only ones surviving the trace are the ones that respect a reflection symmetry along the vertical axis. Such states are fixed by determining the state in half of the volume. The number of these states corresponds to the square root of the total number of states in the whole Hilbert space. By restricting ourselves to specific sectors of the Hilbert space, the two halves are not independent. Nonetheless, we still expect the number of zero-modes to remain exponentially large on the volume.

\section{Low Entropy Zero-Modes in Truncated Link Models}

Here we describe a generic mechanism to obtain low-entropy mid-spectrum states in the TLM, from the vast space of zero-modes.
We start by searching for zero-modes $\ket{\psi_z}$, with a generic decomposition following $\ket{\psi_z} = \sum_n c_i \ket{\phi_i}$, where $\left\{\ket{\phi_i}\right\}$ is the electric field basis. 
By definition, a zero-mode follows
\begin{align}
\label{eq:kinetic-operator-contributions}
    K \ket{\psi_z} = \sum_{i, n, \sigma} c_i H^\sigma_n \ket{\phi_i}
    = 0,
\end{align}
where $\sigma \in \{+, -\}$.
This is very general and must be satisfied by \emph{all} zero-modes.

Spin-$S$ TLMs have an especially simple structure, which we can use to find low-entropy zero-modes. If $H^{\sigma}_{n}\ket{\phi_i}\neq 0$ then $||H^{\sigma}_{n}\ket{\phi_i}|| = 1$ for all $n$, $i$ and $\sigma$. 
This suggests that we can look for solutions where each term in \eqref{eq:kinetic-operator-contributions} is canceled by exactly one other term in the sum. In other words, for each non-null term $c_i H^\sigma_n \ket{\phi_i}$, there exists a term $c_k H^{\sigma'}_{n'} \ket{\phi_k}$ with $c_k = - c_i$ such that $H^{\sigma}_{n}\ket{\phi_n}= H^{\sigma^\prime}_{n^\prime}\ket{\phi_k}$ for some $\sigma', n'$. This introduces the minimal possible number of non-zero weights to cancel a given $H^{\sigma}_{n}\ket{\phi_i}$. 
In TLMs, states $\ket{\phi_i}$, $\ket{\phi_k}$ that obey $H^{\sigma}_{n}\ket{\phi_i}= H^{\sigma^\prime}_{n^\prime}\ket{\phi_k} \neq 0$, are related by $\ket{\phi_k}=H^{-\sigma^\prime}_{n^\prime}H^{\sigma}_{n}\ket{\phi_i}=H^{\sigma}_{n}H^{-\sigma^\prime}_{n^\prime}\ket{\phi_i}$. 

We then focus on two options to cancel a term $c_i H^\sigma_n \ket{\phi_i}$ in Eq.~\eqref{eq:kinetic-operator-contributions} for TLMs:
\begin{enumerate}
    \item $\ket{\phi_i}$ is already annihilated by $H^{\sigma}_{n}$;
    \item The zero-mode also includes the term $-c_i \ket{\phi_k}=-c_i H^{-\sigma^\prime}_{n^\prime}H^{\sigma}_{n}\ket{\phi_i}$ with some $\sigma'$, $n'$.
\end{enumerate}

The second option not only assures that $H^{\sigma}_{n}\ket{\phi_i} - H^{\sigma'}_{n^\prime}\ket{\phi_k} = 0$, but also that $H^{-\sigma'}_{n'}\ket{\phi_i} - H^{-\sigma}_{n}\ket{\phi_k} = 0$.
This means that the contributions $c_i H^{-\sigma^\prime}_{n^\prime}\ket{\phi_i}$ and $c_k H^{-\sigma}_{n}\ket{\phi_k}$ also cancel each other.
What remains are the terms $c_i H^{-\sigma}_{n}\ket{\phi_i}$, $c_i H^{\sigma'}_{n'}\ket{\phi_i}$, $c_k H^{-\sigma'}_{n'}\ket{\phi_k}$ and $c_k H^{\sigma}_{n}\ket{\phi_k}$ as well as all terms that act on different sites.
These also need to be canceled to get a zero-mode.
If it is possible to find a set of states that manage to cancel all contributions in this fashion, this state is a good candidate to be a zero-mode with anomalously low entropy.

Canceling the remaining contributions, could, in principle, be achieved by considering any of these states and acting upon them with another kinetic operator. For example, $H^{\sigma}_{n}\ket{\phi_k}$ could be cancelled by adding a state of the form $H^{\sigma^{\prime\prime}}_{n^{\prime\prime}}H^{\sigma}_{n}\ket{\phi_k}$, for arbitrary choices of $n^{\prime\prime}$ and $\sigma^{\prime\prime}$ that do not annihilate the state. To arrive at the zero-modes in arbitrary integer spin, as described in the main text, we choose to cancel them using the same operators as before, so with the state $\ket{\phi_l} = H^{-\sigma^\prime}_{n^\prime}H^{\sigma}_{n}\ket{\phi_k}$. This cancels two contributions at the same time, $H^{\sigma}_{n}\ket{\phi_k}$ and $H^{-\sigma^\prime}_{n^\prime}\ket{\phi_k}$, further reducing the number of states that need to be included. We then build chains of such states, where both ends are annihilated by the remaining kinetic operators. This way, we build the $2 \times 1$ zero-mode state that can be tiled according to the product defined in the following section.

All analytically constructed zero-mode scars that have been derived for the $S=1/2$ QLM \cite{biswas2022scars, sau2023sublattice} have this structure and therefore an equal magnitude of weights for all contributing electric field basis states.

The outlined strategy can be used to construct more zero-mode scars in systems that are not accessible through ED. 
For other formulations, like QLMs, it will still be true that states that cancel each other in \eqref{eq:kinetic-operator-contributions} differ by two applications of a kinetic operator. This insight was crucial to understanding the low entropy states observed here for TLMs and may prove useful in the analysis of other systems.

\section{The tiling product}

There are multiple options for generalizing scars found in small systems to arbitrarily large lattices. The most straightforward approach is to apply tensor products in the electric field basis. This is only possible if the links at the boundary agree and the resulting system still follows Gauss' law. 
We generalize this to the \emph{tiling product} $\odot$, which acts like a tensor product on the dual representation instead. 
It is defined as
\begin{equation}
    \ket{M} \odot \ket{M^\prime} = \ket{M M^\prime},
\end{equation}
i.e. the matrices of height variables are concatenated. This is not the tensor product, since the values of the boundary links are added. If the concatenation of $M$ and $M^\prime$ would give rise to an invalid gauge field configuration, i.e. some neighboring height variables satisfy $|h_i - h_j| > S$, we will say that $\ket{M}$ and $\ket{M'}$ \emph{do not tile} and define their tilling product to give the null state. 
More generally, we are interested in tiling products of the form $\left(\sum_ac_a\ket{M_a}\right)\odot\left(\sum_bc_b\ket{M^\prime_b}\right)$. We set this product to result in the null state if \emph{any} state in the left-hand side does not tile with \emph{any} state in the right side, otherwise, we just apply the distributive law. In other words, the tiling is distributive if \emph{all} the states in one entry tile with \emph{all} the states in the other.

For example, states that only contain height variables $h_i$ with $\max_i(h_i)-\min_i(h_i) \leq S$, will always tile with themselves. If the difference is larger, one might need to be careful which plaquettes can touch, to prevent neighboring height variables from differing by more than $S$.

This tiling product gives non-null states (specifically the scars) for the tiles constructed in the main text since $\max_i(h_i)-\min_i(h_i) = S$. In general, scars do not need to have a tiling structure. One eigenstate of $E^2$ is not composed of tilings. It is described in the following section.

\section{A $4\times 4$ Scar for the $E^2$ potential}

\begin{figure}
    \centering
    \includegraphics[width=.45\linewidth]{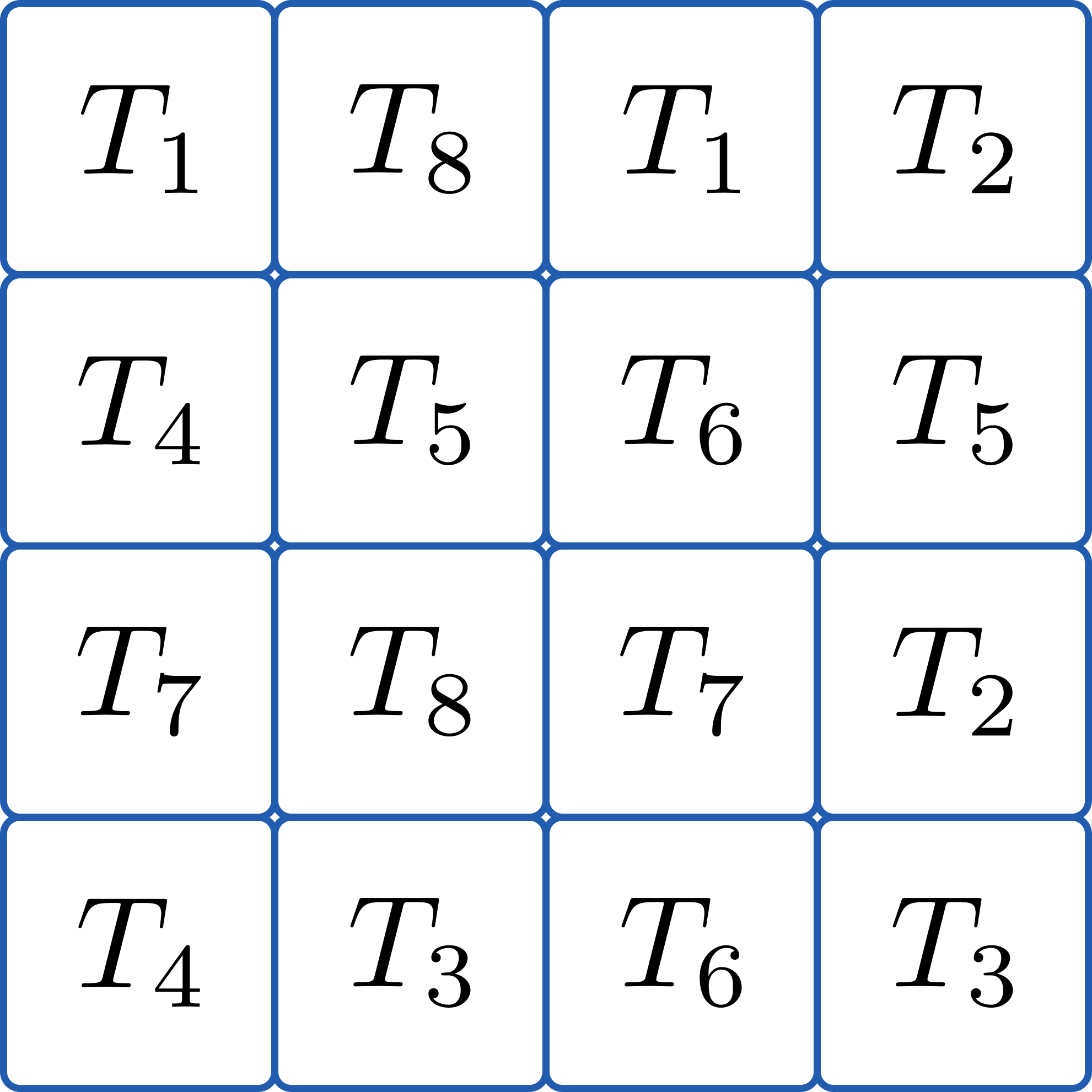}
    \caption{Schematic representation of a tiling that produces a scar for the $E^2$ potential in a $4\times 4$ lattice.}
    \label{fig:exclusive4x4}
\end{figure}

The remaining observed scar referred to in the main text seems exclusive of the $4\times 4$ system. While the fundamental tiling structure still applies, it comes from a different tiling of the states described in the main text. The two plaquettes in a tile lie in the same row or column but have a distance of 2. Explicitly

\begin{equation}
\label{eq:full-scar-states}
    \ket{\psi_s} = \frac{1}{16}\prod_{(n,n') \in T}
    \left(H^+_n-H^+_{n'}\right) \ket{\mathbf{0}},
\end{equation}
where the set of tiles $T=\left\{T_1,\dots,T_8\right\}$ is presented in Fig. \ref{fig:exclusive4x4}.
This tiling guarantees that each $T_i$ shares the same number of neighboring height variables of value 0 and 1 and therefore will be an eigenstate of the potential with eigenvalue $16g$.

\section{Amplitudes of Scars}

The amplitudes of the isolated low-entropy states match the predictions of the constructed scars. We demonstrate this in the case of the negative energy scars in $S=1$ single-leg ladders.

The predicted scars with negative energy are generated by tilings of $\ket{\psi_-}$ defined in the main text. In the single-leg ladder, only two tilings exist. This generates two states of equal energy, which we call $\ket{\psi_1}$ and $\ket{\psi_2}$. They are linearly independent and will be found in a superposition $\ket{\psi_S} = a \ket{\psi_1} + b\ket{\psi_2}$. All non-zero amplitudes $\braket{\phi_{i}|\psi_{1,2}}$, in the electric field basis, are $\pm \frac{1}{\sqrt{2^{L/2}}}$. 
\begin{figure}[h]
    \centering
    \includegraphics[width=0.7\linewidth]{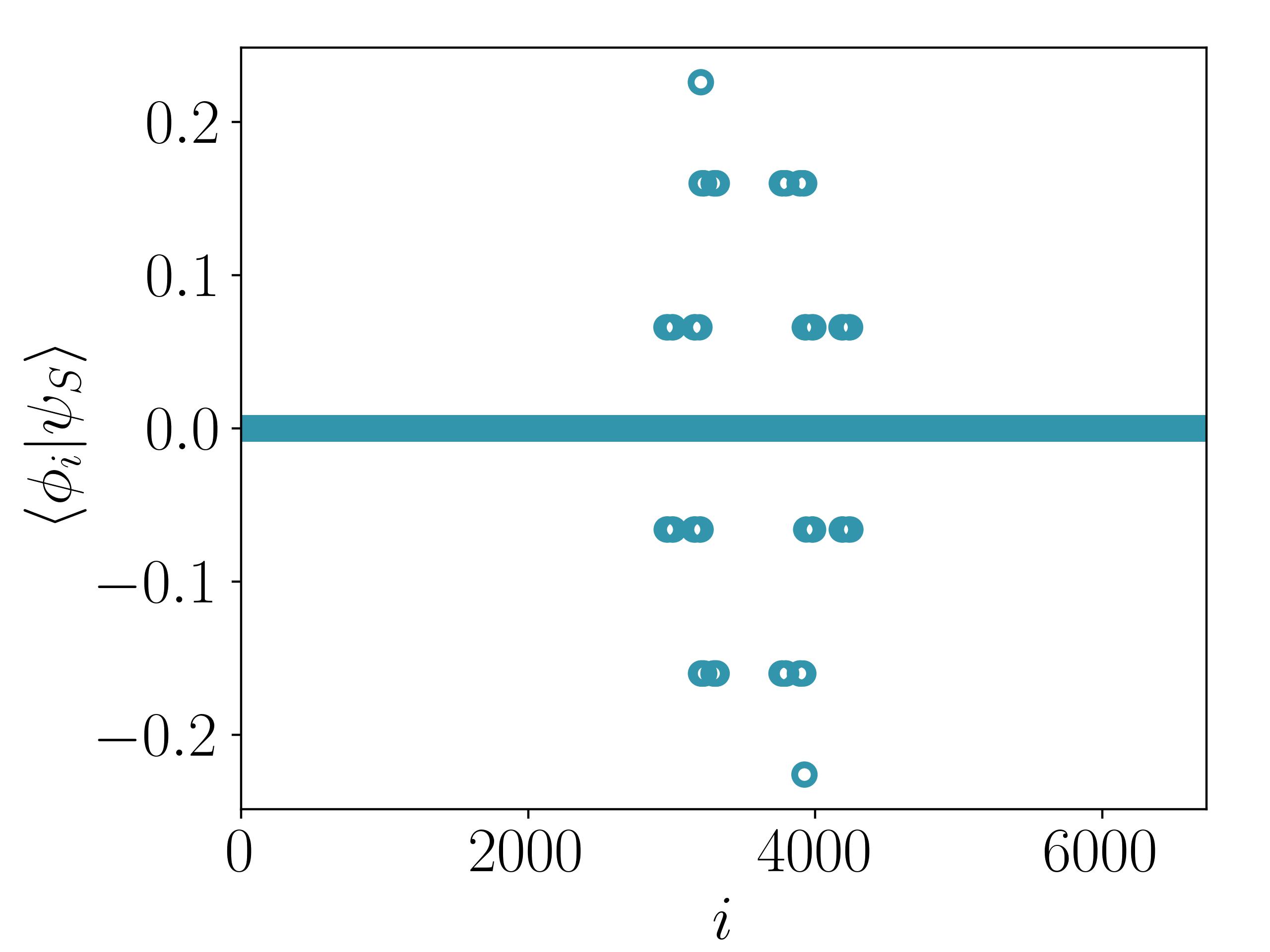}
    \caption{Overlap of electric field basis states $\ket{\phi_i}$ with the lowest entropy scar in Fig.~3 of the main text. It depicts the entanglement entropy of $10 \times 1$ $S=1$ plaquettes with periodic boundary conditions in the $x$ direction. Most states have zero overlap.
    }
    \label{fig:overlap}
\end{figure}
There are therefore four different possible magnitudes of the amplitudes $A = \braket{\phi_i|\psi_S}$:
\begin{enumerate}
    \item The basis state $\ket{\phi_i}$ has zero overlap with both $\ket{\psi_1}$ and $\ket{\psi_2}$ and therefore the amplitude is zero.
    \item The basis state has non-zero overlap with only $\ket{\psi_1}$ and has the amplitude $A  = \pm \frac{a}{\sqrt{2^{L/2}}}$.
    \item The basis state has non-zero overlap with only $\ket{\psi_2}$ and has the amplitude $A  = \pm \frac{b}{\sqrt{2^{L/2}}}$.
    \item The basis state has non-zero overlap with both $\ket{\psi_1}$ and $\ket{\psi_2}$ and has the amplitude $A  = \pm \frac{a+b}{\sqrt{2^{L/2}}}$.
\end{enumerate}
The fourth case only occurs for the states $\ket{10101010...}$ and $\ket{01010101...}$, which exist in both tilings.
The eight different amplitudes are visible in Fig.~\ref{fig:overlap}. It shows the overlap of the electric field basis states $\ket{\phi_i}$ with a scar of a $10 \times 1$ ladder, with spin $S=1$, under periodic boundary conditions in the $x$ direction.

\end{document}